\documentclass[aps,prc,reprint,nofootinbib]{revtex4-1}
\usepackage{graphicx,graphics}   %       for graphics
\usepackage{latexsym}   %       for special symbols
\usepackage{mathrsfs}
\usepackage{fancyhdr}
\usepackage{amssymb}
\usepackage{amsfonts}
\usepackage{setspace}
\usepackage{float}
\usepackage{amsmath}
\usepackage{bm}

\usepackage{natbib,twoopt}
\usepackage{hyperref}
\hypersetup{
    colorlinks=true,
    linkcolor=red,
    citecolor=blue,
    filecolor=magenta,      
    urlcolor=cyan,
    bookmarks=true,
}  %% to set colour pattern for hyperref
\begin{document}
%==============================================================================

%==============================================================================
\title{Higher-order baryon number susceptibilities:\\ interplay between the chiral and the nuclear liquid-gas transitions}

\author{A. Mukherjee$^{1,2}$, J. Steinheimer$^1$ and S. Schramm$^{1,2}$}

\affiliation{$^1$ Frankfurt Institute for Advanced Studies, Ruth-Moufang-Stra{\ss}e 1, D-60438 Frankfurt am Main, Germany}
\affiliation{$^2$ Institut f\"ur Theoretische Physik, Goethe Universit\"at Frankfurt, Max-von-Laue-Stra{\ss}e 1, D-60438 Frankfurt am Main, Germany}

%\date{February 28, 2014}

\begin{abstract}
We use an improved version of the SU(3) flavour parity-doublet quark-hadron model to investigate the higher order baryon number susceptibilities near the chiral and the nuclear liquid-gas transitions. The parity-doublet model has been improved by adding higher-order interaction terms of the scalar fields in the effective mean field Lagrangian, resulting in a much-improved description of nuclear ground-state properties, in particular the nuclear compressibility. The resulting phase diagram of the model agrees qualitatively with expectations from lattice QCD, i.e., it shows a crossover at zero net baryo-chemical potential and a critical point at finite density. Using this model, we investigate the dependence of the higher-order baryon number susceptibilities as function of temperature and chemical potential. We observe a strong interplay between the chiral and liquid-gas transition at intermediate baryo chemical potentials. Due to this interplay between the chiral and the nuclear liquid-gas transitions, the experimentally measured cumulants of the net baryon number may show very different beam energy dependence,
subject to the actual freeze-out temperature.
\end{abstract}

%\pacs{}

\maketitle
\section{Introduction}
The theory of quantum chromodynamics (QCD) is expected to have a rich phase structure at finite chemical potential and temperature \cite{Stephanov:1998dy,McLerran:2007qj,Alford:2007xm}. Its study is a central topic of high energy nuclear physics. Experimental programs at the Large Hadron Collider (LHC) and the Relativistic Heavy Ion Collider (RHIC) are currently investigating the properties of hot and dense QCD matter. Future programs at RHIC, the Facility for Anti-Proton and Ion Research (FAIR) and the NICA facility are aimed at a better understanding of the phase transition from hadronic to de-confined quark matter and the transition from a phase where chiral symmetry is broken to one where it is restored.\\
Theoretical studies employing lattice QCD methods have already established that the transition from hadrons to quarks proceeds as a smooth crossover in the case of vanishing net baryon number density \cite{Borsanyi:2010cj,Bazavov:2010sb}. For finite net baryon density, the use of standard lattice QCD methods is limited by the so-called fermion sign problem. Some conclusions can be drawn by extending the lattice thermodynamic quantities, via a Taylor expansion around $\mu_\text{B}=0$ \cite{Allton:2002zi,Gunther:2016vcp}, for values of $\mu_{\rm B}/T<2$. However, to go to even higher densities, higher orders of the expansion coefficients need to be known to a very good accuracy, a requirement which cannot be met with the current computational possibilities. Thus, the current conclusions are that a first order phase transition seems very unlikely for chemical potentials smaller than $\mu_\text{B}/T \approx 2-3$ (i.e., for $\mu_{\rm B} < 200-300$ MeV).\\
At very large net baryon densities and low temperatures, astrophysical observations may also help to constrain the QCD Equation of State. Nuclear matter ground-state properties have been derived from measurements to a high accuracy. On the other hand, properties of compact stars like their masses and eventually, their radii, can serve as important information for determining the equation of state at several times nuclear ground-state density (c.f., e.g., \cite{Steiner:2010fz}).\\
\noindent
With the currently available information, we can theorise that, in the temperature ($T$) and baryo-chemical potential ($\mu_\text{B}$) plane, the conjectured QCD phase diagram consists, primarily, of three parts: 
\begin{enumerate}
\item a high $\mu_\text{B}$ and temperature region, where chiral symmetry is restored, containing a de-confined quark-gluon plasma (QGP),
\item a low temperature and moderate $\mu_\text{B}$ region containing dense, strongly interacting nuclear matter and nuclei, and
\item a low temperature and low chemical potential region made up of purely hadronic matter described by a hadron resonance gas.
\end{enumerate}
The transition from a phase where chiral symmetry is broken to one where it is restored, for lower temperatures, is conjectured to be a first-order phase transition, which switches to a continuous transition at a point known as the 'Critical Point' ($T_\text{C}$). At values of $\mu_\text{B}$ lower than that at $T_\text{C}$, the transition is termed a 'Crossover Transition'. The transition from a dilute gas of hadrons to bound nuclear matter is also a first-order phase transition, generally called the nuclear liquid-gas phase transition.\\
\noindent
In order to verify this suggested phase structure experimentally, heavy-ion collision experiments are conducted at various beam energies. Some particularly interesting observables in these experiments are the moments of the multiplicity distributions, the susceptibilities of the different conserved net charges (baryon number, strangeness and electric charge) \cite{Aggarwal:2010wy,PhysRevLett.112.032302}. The reason behind this interest is that one proposed universal characteristic of the critical point of QCD is the divergence of the correlation length, $\xi \rightarrow \infty$, of the order-parameter ($\sigma$ and $\zeta$) fields. As a consequence, higher-order fluctuation moments of observables coupled to these fields also diverge, at least for an infinite system size and relaxation time \cite{Stephanov:2008qz,Stephanov:1998dy}.\\
\noindent
The direct comparison of these measured susceptibilities with lattice QCD results is neither straightforward, nor entirely unambiguous. Currently known complications in the interpretation of such measurements include: corrections of experimental efficiency and acceptance effects \cite{Bzdak:2012ab,Bzdak:2016qdc,kitazawa2016efficient}, cluster formation \cite{Feckova:2015qza}, conservation laws \cite{Begun:2004gs,Bzdak:2012an}, corrections due to the finite size of the system \cite{Gorenstein:2008et}, fluctuations of the system volume\cite{Gorenstein:2011vq,Sangaline:2015bma} and fluctuations present in the initial state of the collision \cite{Spieles:1996is}. Moreover, it has been pointed out that the hadronic decoupling phase, which occurs after hadronisation, can have a strong influence on the observed multiplicity distributions \cite{Kitazawa:2012at,Asakawa:2000wh,Jeon:2000wg,Steinheimer:2016cir}.\\
On the theoretical side, the measured susceptibilities, at least for large values of the chemical potential, are usually compared to effective models, which normally do not include a de-confinement or chiral transition and a nuclear liquid-gas transition simultaneously.

\noindent
Current heavy-ion experiments create systems with widely varying $\mu_{\rm B}$ and high $T$ values. In fact only the experiments at the LHC and high-energy RHIC can be directly connected to LQCD results. Most interesting results on the baryon number susceptibilities, which is the central 
topic of the manuscript, have been obtained at rather low beam energies e.g. at the RHIC beam energy scan and the HADES experiment at GSI. Here the values of chemical potential are large and temperatures are moderate.
In our paper we focus on the fact that at such high values of chemical potential ($\mu_{\rm B} > 400$ MeV) the effect of nuclear interactions can and should not be neglected.
In this paper, we will discuss how the observed susceptibilities may change if one takes into account an equation of state that includes a nuclear liquid-gas transition, as well as a first order chiral transition at high baryon densities. We will also show, how the observed susceptibilities change with beam energy for different freeze-out lines in the phase diagram and how the interplay between the liquid-gas transition and the chiral transition manifests itself in the beam-energy dependence.

\section{Model Description}
\subsection{The Parity-Doublet Model}

The parity-doublet model, as used in this paper, serves as an effective approach to describe the strongly interacting hadronic, and in extension, quark matter.
In this approach, an explicit mass term for baryons in the Lagrangian is possible, which preserves chiral symmetry. Here, the signature for chiral symmetry restoration is the degeneracy of the usual baryons and their respective negative-parity partner states.
In the following, we outline the basic SU(3) parity model and determine nuclear matter saturation properties with this ansatz. Subsequently, we calculate the phase diagram of isospin-symmetric
matter by varying the
baryo-chemical potential and temperature of the system. 

\noindent
In the model approach, positive and negative parity states of the baryons
are grouped in doublets $N = (N^+,N^-)$ as discussed in  \cite{PhysRevD.39.2805,Hatsuda:1988mv}.
The flavour SU(3) extension of the approach, using the non-linear representation of the fields, is quite straightforward, as shown in \cite{Nemoto:1998um} and details can be found in \cite{Steinheimer:2011ea}.
In addition, as outlined in \cite{Papazoglou:1997uw}, one constructs SU(3)-invariant terms in the Lagrangian including the meson-baryon and meson-meson self-interaction terms.

\noindent
Taking into account the scalar and vector condensates in mean-field approximation, the resulting Lagrangian ${\cal L_{\rm B}}$ reads as \cite{Steinheimer:2011ea}
\begin{eqnarray}
{\cal L_{\rm B}} &=& \sum_i (\bar{B_\text{i}} i {\partial\!\!\!/} B_\text{i})
+ \sum_i  \left(\bar{B_\text{i}} m^*_\text{i} B_\text{i} \right) \nonumber \\ &+&
\sum_i  \left(\bar{B_\text{i}} \gamma_\mu (g_{\omega \text{i}} \omega^\mu +
g_{\rho \text{i}} \rho^\mu + g_{\phi \text{i}} \phi^\mu) B_\text{i} \right) ~,
\label{lagrangian2}
\end{eqnarray}
summing over the states of the baryon octet. The effective masses of the baryons (assuming isospin-symmetric matter) are
\begin{eqnarray}
m^*_{\text{i}\pm} = \sqrt{ \left[ (g^{(1)}_{\sigma \text{i}} \sigma + g^{(1)}_{\zeta \text{i}}  \zeta )^2 + (m_0+n_\text{s} m_\text{s})^2 \right]}\nonumber \\
\pm g^{(2)}_{\sigma \text{i}} \sigma \pm g^{(2)}_{\zeta \text{i}} \zeta ~,
\label{effmass}
\end{eqnarray}
with the $g^\text{(j)}_\text{i}$ as the coupling constants of the baryons with the scalar fields. In addition, there is an SU(3) symmetry-breaking mass term proportional to the strangeness, $n_\text{s}$, of the respective baryon.
Note that the parity-doublet models allow for two different scalar coupling terms $i=1,2$.

\noindent
The scalar meson interaction, driving the spontaneous breaking of the chiral symmetry,
can be written in terms of SU(3) invariants
$I_2 = (\sigma^2+\zeta^2) ,~ I_4 = -(\sigma^4/2+\zeta^4) $ and $I_6 = (\sigma^6 + 4\zeta^6) $ as:
\begin{equation}
V = V_0 + \frac{1}{2} k_0 I_2 - k_1 I_2^2 - k_2 I_4 + k_6 I_6 ~,
\label{veff}
\end{equation}
where $V_0$ is fixed by demanding a vanishing potential in the vacuum.\\ 
In this work, the last term, $k_6 I_6$, has been introduced following Ref. \cite{Motohiro:2015taa}, which results in an improved lowering of the calculated nuclear matter compressibility value to 267 MeV, which is now in reasonable agreement with the phenomenologically obtained range of about $200-280$ MeV.

\noindent
The set of scalar coupling constants are fitted in order to reproduce the vacuum masses of the nucleon, and the
$\Lambda$, $\Sigma$, and $\Xi$ hyperons , whereas the vector couplings are chosen to reproduce reasonable values for nuclear ground-state
properties (see Ref. \cite{Steinheimer:2011ea}).\\
\noindent
As a likely parity partner of the nucleon we choose the N(1535) resonance with its correspondent mass.
In order to keep the number of parameters  small, we assume equal mass splitting of the baryons with their respective parity partners, therefore setting $g^{(2)}_{\zeta \text{i}} = 0$. 
An SU(3) description, in addition to enhancing the number of degrees of freedom, also necessarily increases the
number of parameters. In order to avoid being overwhelmed by too many new parameters, we assume, for simplicity, that the splitting of the
various baryon species and their respective parity partners is of the same value for all baryons, which is achieved by setting
$g^{(2)}_{\sigma \text{i}} \equiv g^{(2)}$ and $g^{(2)}_{\zeta \text{i}} = 0$.\\
\noindent
The hyperonic vector interactions were tuned to generate phenomenologically  acceptable
optical potentials of the hyperons in ground-state nuclear matter, with $U_\Lambda(\rho_o) = -28\,{\rm MeV}$ and $U_\Xi(\rho_o) =  -18\,{\rm MeV}$. The mass difference due to the strange quark was fixed at $m_\text{s} = 150$ MeV. All parameter values used are summarized in Table \ref{modpar}.

\begin{table}[t]
\begin{center}
\begin{tabular}{ | c | c | c | }
\hline
 $k_0$ & $k_1$ & $k_2$ \\ 
 $(242.61 \text{ MeV})^2$ & 4.818 & -23.357 \\
 \hline
 $k_6$ & $\epsilon$ & $g_\sigma^1$ \\
 $(0.276)^6 \text{ MeV}^{-2}$ & $(75.98 \text{ MeV})^4$ & \hspace{1.5mm} -8.239296 \hspace{1.5mm} \\
 \hline
 $g_\sigma^8$ & $\alpha_\sigma$ & $g_{\text{N}\omega}$ \\ 
 -0.936200 & 2.435059 & 5.45\\
 \hline
\end{tabular}
\caption{Model Parameters: the F, D, and S-type couplings $\alpha_\sigma$, $g_\sigma^8$ and $g_\sigma^1$ determine the couplings of the various baryons.}
\label{modpar}
\end{center}
\end{table}

\subsection{Mesons and Quarks}
At some temperature, QCD exhibits a transition from a hadronic to a de-confined phase, at which point, the quarks become the dominant degrees of freedom. This transition occurs as a smooth crossover, at least for $\mu_{\mathrm{B}}=0$. Consequently there has been discussion about the actual temperature up to which a hadronic description is still valid \cite{Aoki:2006we,Bazavov:2010bx,Vovchenko:2015cbk}. We can only say for sure that the order parameter of the chiral transition, the chiral condensate, has an inflection point at the pseudo-critical temperature $T_{\mathrm{PS}} \approx 155$ MeV, and that de-confinement occurs in a temperature region of $T_\text{dec}\approx 150 - 400$ MeV. Nevertheless, at some point, the hadronic parity-doublet model will not be the appropriate effective description of QCD and one needs to introduce a de-confinement mechanism in the model. In this work, we will apply a mechanism that has been introduced in \cite{Steinheimer:2010ib}, to add a de-confinement transition in a chiral hadronic model. This is done by adding an effective quark and gluon contribution, as done in the PNJL approach \cite{Fukushima:2003fw,Ratti:2005jh}. This model uses the Polyakov loop $\Phi$ as the order parameter for de-confinement. $\Phi$ is defined via
\begin{equation}
\Phi=\frac{1}{3}\text{Tr}[\exp{(i\int d\tau A_4)}],
\label{phidef}
\end{equation}
where $A_4=iA_0$ is the temporal component
of the SU(3) gauge field, distinguishing $\Phi$ and its conjugate $\Phi^{*}$, at finite baryon densities \cite{Fukushima:2006uv,Allton:2002zi,Dumitru:2005ng}.\\
The effective masses of the quarks
are generated by the scalar mesons, except for a small explicit
mass term ($\delta m_\text{q}=5$ MeV and $\delta m_\text{s}=150$ MeV, for the strange quark) and $m_0$:
\begin{eqnarray}
&m_\text{q}^*=g_{\text{q}\sigma}\sigma+\delta m_\text{q} + m_{0\text{q}}&\nonumber\\
&m_\text{s}^*=g_{\text{s}\zeta}\zeta+\delta m_\text{s} + m_{0\text{q}},&
\end{eqnarray}
with values of $g_{\text{q}\sigma}=g_{\text{s}\zeta}= 4.0$. Similar to the case of the baryons, we also introduced a mass parameter $m_{0\text{q}}= 165$ MeV for the quarks. Again, this additional mass term can be due to a coupling of the quarks to the dilaton field (gluon condensate). Given such a mass term, the quarks do not appear in the nuclear ground-state, which would be a clearly non-physical result. This also permits us to set the vector type repulsive interaction strength of the quarks, to zero. A non-zero vector interaction strength would lead to a massive deviation of the quark number susceptibilities from lattice data, as has been indicated in different mean field studies \cite{kunihiro1991quark,Ferroni:2010xf,Steinheimer:2010sp,steinheimer2014lattice}.\\
A coupling of the quarks to the Polyakov loop is introduced in the thermal energy of the quarks. Their thermal contribution to the grand-canonical potential $\Omega$, can be written as:
\begin{equation}
\Omega_\text{q}=-T \sum_{{\rm i}\in Q}{\frac{\gamma_\text{i}}{(2 \pi)^3}\int{d^3k \ln\left(1+\Phi \exp{\frac{E_\text{i}^*-\mu_\text{i}}{T}}\right)}}
\end{equation}
and
\begin{equation}
\noindent
\Omega_{\overline{\text{q}}}=-T \sum_{\text{i}\in Q}{\frac{\gamma_\text{i}}{(2 \pi)^3}\int{d^3k \ln\left(1+\Phi^* \exp{\frac{E_\text{i}^*+\mu_\text{i}}{T}}\right)}}~.
\end{equation}
The sums run over all quark flavours, where $\gamma_\text{i}$ is the corresponding degeneracy factor, $E_\text{i}^*=\sqrt{m_\text{i}^{*2}+p^2}$ the energy and $\mu_\text{i}^*$ the chemical potential of the quark.\\
All thermodynamic quantities: energy density $e$, entropy density $s$, as well as the
densities of the different particle species $\rho_\text{i}$, are derived from the grand-canonical potential.
It includes the effective potential $U(\Phi,\Phi^*,T)$, which controls the dynamics of the Polyakov loop.
For simplicity, in our approach we adopt the ansatz proposed in \cite{Ratti:2005jh}:
\begin{eqnarray}
	U&=&-\frac12 a(T)\Phi\Phi^*\nonumber\\
	&+&b(T)\ln[1-6\Phi\Phi^*+4(\Phi^3\Phi^{*3})-3(\Phi\Phi^*)^2],
\end{eqnarray}
 with $a(T)=a_0 T^4+a_1 T_0 T^3+a_2 T_0^2 T^2$, $b(T)=b_3 T_0^3 T$.\\
The parameters $a_0, a_1, a_2$ and $b_3$ are initially fixed, as in \cite{Ratti:2005jh}, by demanding a first order phase transition in the pure gauge sector at $T_0=270$ MeV, and that the Stefan-Boltzmann limit of a gas of gluons is reached for $T\rightarrow\infty$. In general, of course, the presence of quarks may have a significant influence on the Polyakov potential \cite{Schaefer:2007pw}, and in order to obtain a crossover transition at $\mu_\text{B}=0$, we change $T_0$ to 200 MeV.\\
\noindent
In the following, as a way to remove the hadrons once quarks are de-confined, we introduce excluded volumes for the hadrons in the system.
Including effects of finite-volume particles in a thermodynamic model for hadronic matter was proposed long ago \cite{Hagedorn:1980kb,Baacke:1976jv,Gorenstein:1981fa,Hagedorn:1982qh,Rischke:1991ke,Cleymans:1992jz,Kapusta:1982qd,Bugaev:2000wz,Bugaev:2006pt,Satarov:2009zx}. In recent publications \cite{Steinheimer:2010ib,Steinheimer:2010sp}, we adopted this ansatz to successfully describe a smooth transition from a hadronic to a quark dominated system (see also \cite{Sakai:2011fa}).\\
In particular, we introduce the quantity $v_\text{i}$, which is the volume excluded by a particle of species $i$, where we only distinguish between baryons, mesons and quarks. Consequently, $v_\text{i}$ can assume three values:
\begin{eqnarray}
 v_\text{quark}&=&0 \nonumber \\
 v_\text{baryon}&=&v \nonumber \\
 v_\text{meson}&=&v/a, \nonumber
\end{eqnarray}
where $a$ is a number larger than one. In our calculations, we choose a value of $a=8$, which assumes that the radius $r$ of a meson is half of the radius of a baryon.
Note that at this point, we neglect any possible density-dependent and Lorentz contraction effects on the excluded volumes as introduced in \cite{Bugaev:2000wz,Bugaev:2006pt}.
The modified chemical potential $\widetilde{\mu}_\text{i}$, which is connected to the real chemical potential $\mu_\text{i}$ of the $i$-th particle species, is obtained by the following relation:
\begin{equation}
	\widetilde{\mu}_\text{i}=\mu_\text{i}-v_\text{i} \ P,
\end{equation}
where $P$ is the sum over all partial pressures. To be thermodynamically consistent, all densities ($\widetilde{e_\text{i}}$, $\widetilde{\rho_\text{i}}$ and $\widetilde{s_\text{i}}$) have to be multiplied by a volume correction factor $f$, which is the ratio of the total volume $V$ and the reduced volume $V'$, not being occupied:
\begin{equation}
	f=\frac{V'}{V}=(1+\sum_{\rm i}v_\text{i}\rho_\text{i})^{-1}
\end{equation}
\begin{equation}
  e=\sum_{\rm i} f \ \widetilde{e_\text{i}}, \quad
  \rho_\text{i}=f \ \widetilde{\rho_\text{i}}, \quad
  s=\sum_{\rm i} f \ \widetilde{s_\text{i}}~.
\end{equation}
As a consequence, the chemical potentials of the hadrons are decreased by the quarks, but not vice-versa. In other words, as the quarks start appearing, they effectively suppress the hadrons by changing their chemical potential, while the quarks are only affected through the volume correction factor $f$.

\section{Results}

\subsection{Comparison with Lattice QCD}
\label{latsec}
\begin{figure}[t]
\includegraphics[width=0.35\textwidth, angle = 270]{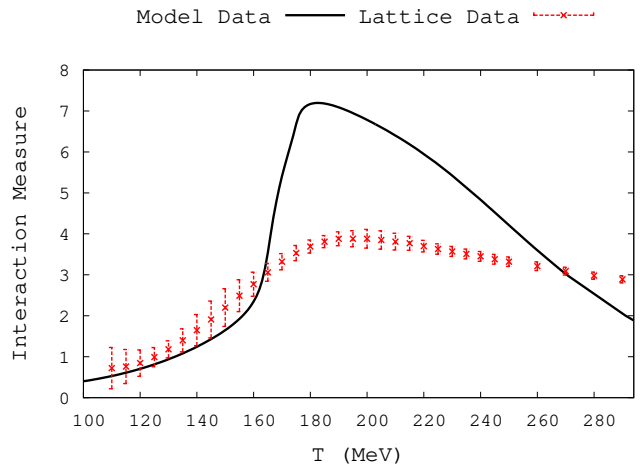}
\caption{[colour online] Interaction measure, from the model (at $\mu_\text{B}$ = 0) and lattice data, as a function of temperature.}
\label{ept}
\end{figure}

A comparison with data obtained from lattice QCD calculations is necessary in order to benchmark the model results and their subsequent modifications. To that end, we determine the interaction measure $I$, defined as: 
\begin{equation}
I = \frac{\varepsilon-3P}{T^4} ~,
\label{intmes}
\end{equation}
with $\varepsilon, P$ and $T$ as the energy density, pressure and temperature, respectively. The model result for $I$ at $\mu_\text{B}=0$ as function of temperature, in comparison to available lattice data \cite{Borsanyi:2013bia}, is shown in Fig. \ref{ept}.
We observe that, indeed, the model gives a good description of
LQCD thermodynamics below the pseudo-critical temperature $T_{\rm C}$. But, although the shapes obtained from both sets of data are similar, the peak value of the interaction measure is much higher in case of the parity-doublet model than that obtained from lattice QCD. This is likely a result of our use of the standard Polyakov loop potential for the description of the quark and gluon deconfinement. For future investigations, it is therefore interesting to implement an improved version of the Polyakov potential which better describes the thermodynamics at $\mu_{\rm B}=0$. At this point, we want to clarify the intent of this paper again: instead of constraining our model parameters by a fit to lattice QCD results at $\mu_{\rm B}=0$, we constrain our model parameters by actual observables at large baryon densities and low temperatures, e.g., nuclear ground-state properties and neutron star observations. Starting from these parameters, we then extend the model to low densities where the remaining free parameters (mainly those of the Polyakov loop) are subsequently used, to get at least a reasonable description of low$-\mu_{\rm B}$ lattice results. Within the current set up of the model, a perfect description of lattice QCD results appears to be unachievable.

\noindent
Checking model results for nuclear model ground-state properties, we  obtain  phenomenologically acceptable values of a nucleonic binding energy of $-16.00$ MeV and a compressibility ($\kappa$) of 267.12 MeV, for a saturation density $\rho_0 = 0.142 \text{ fm}^{-3}$. Note that the compressibility, which in general tends to be very high in parity-doublet models, has a reasonable value  (see also \cite{Motohiro:2015taa}).

\subsection{Pressure and Quark Fraction}
\label{pqf}
Some interesting characteristics of the system may be revealed by observing the pressure of the system, along the transition lines, as a function of temperature.

\begin{figure}[t]
\centering
\includegraphics[width=0.35\textwidth, angle = 270]{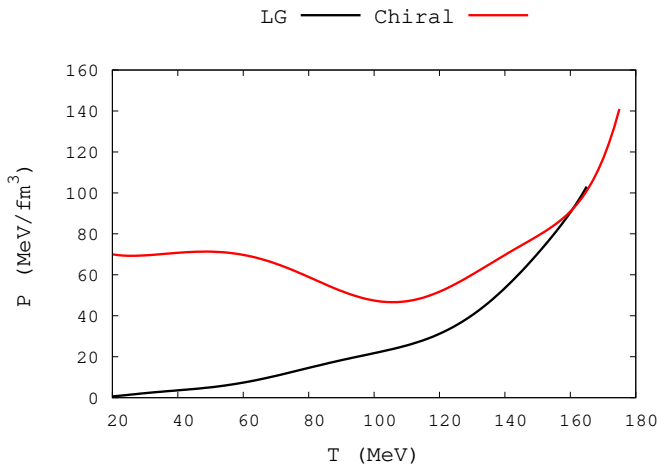}
\caption{[colour online] Pressure, as a function of temperature, along the transition lines.}
\label{pplot}
\end{figure}

\noindent
For the nuclear liquid-gas transition we define this line 
as the maximum of the derivative of the net-baryon density with respect to 
the baryo-chemical potential. Similarly, for the chiral transition, we define it as the maximum of the derivative of the $\sigma$ field (chiral condensate) with respect to the baryo-chemical potential (or the temperature, for baryo-chemical potential values of $\mu_{\rm B} < 400$ MeV, i.e. beyond the merger of the two transition lines). Note that both criteria can be used equivalently for either transition, as the net-baryon density and value of the sigma field are intimately related (c.f. \cite{Walecka:1974qa,Bender:2003jk,Cohen:1991nk} and references therein).
This means that when we observe a rapid change in the net baryon number density, we will also observe a rapid change in the chiral condensate and vice-versa.
Thus, both criteria can be used to identify the crossover lines of the chiral and LG transition.
Note that, if there was an additional separation of the chiral and deconfinement line (e.g., as discussed in \cite{Ferreira:2013tba}), the situation we try to describe would be even more complicated.

\noindent
In the region where both first-order transitions switch to crossovers, we fit a double-Gaussian function to the derivative of the net-baryon density with respect to the baryo-chemical potential, assigning each peak to one transition line. One should, of course, note that the two transitions show clear differences. Even though the value of the chiral condensate changes slightly at the liquid gas transition, chiral symmetry is restored much later; after the chiral transition; 
where the chiral condensate, essentially, drops to zero.

\noindent
The model results for the pressure along the transition and crossover lines, are shown in Fig. \ref{pplot}, where the baryo-chemical potential increases with decreasing temperature along both lines (cf. Fig. \ref{crit}). 
The behaviour of the pressure along the transition line for the liquid-gas transition is as expected. Since the baryon density along the liquid-gas transition does not change considerably with increasing temperature, the change in the pressure is driven, primarily, by the increase in the entropy caused by the increasing temperature. Such a behaviour can be observed when the specific entropy in the gas phase is larger than that in the liquid phase, as derived from the Clapeyron equation \cite{Hempel:2013tfa}. 

\begin{figure}[t]
\centering
\includegraphics[width=0.35\textwidth, angle = 270]{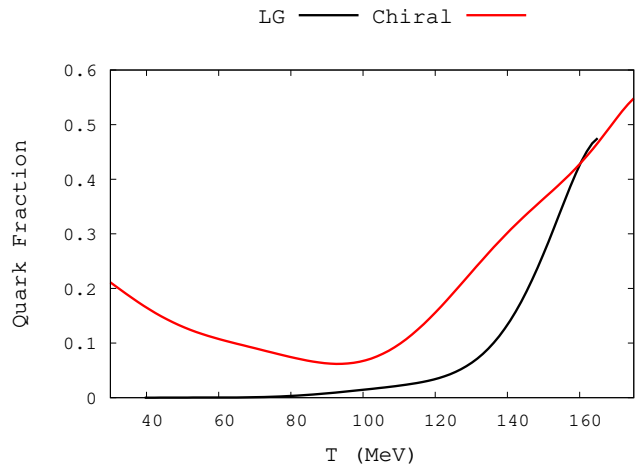}
\caption{[colour online] Quark fraction $q_{\rm f}$, as function of temperature, along the transition lines.}
\label{qfrac}
\end{figure}

\noindent
For the chiral transition, the change of the pressure along the transition line is more complicated. At large temperatures and small chemical potentials, the pressure essentially follows the trend of the nuclear liquid-gas transition, as the meson-dominated system transitions smoothly into a system dominated by quark and gluons. As the chemical potential increases, however, the change in degrees of freedom is manifested more strongly in a change of net baryon number, as the system transitions from heavy baryons to light baryons. Consequently, the change in net baryon number dominates the change of pressure and thus, the pressure along the transition line shows a behaviour opposite to that observed during the liquid-gas transition. As the transition line goes to even lower temperatures, the behaviour of the pressure changes direction again. This time, however, it is a result of the change in curvature of the transition line in the $T$-$\mu_{\text{B}}$ diagram (cf. Fig. \ref{crit}). This is, most likely, an artefact of the Polyakov model, which is not very reliable at low values of temperature and large values of baryo-chemical potential. 
In any case, it is important to note that the pressure at zero temperature for the de-confinement transition takes a finite value, which is an important property of a ``realistic" model for the QCD EoS.

In order to illustrate the change in degrees of freedom at the transition lines, one can determine the so-called quark fraction $q_{\rm f}$, defined as:
\begin{equation}
q_{\rm f} = \frac{\varepsilon_{\text{quark}}+\varepsilon_{\text{Polyakov}}}{\varepsilon_{\text{baryon}}+\varepsilon_{\text{meson}}+\varepsilon_{\text{Polyakov}}}~;
\label{qf}
\end{equation}
with $\varepsilon_{\text{quark}}$, $\varepsilon_{\text{baryon}}$, $\varepsilon_{\text{meson}}$ and $\varepsilon_{\text{Polyakov}}$ denoting the energy density contributions from the quarks, baryons (including quarks), mesons and the Polyakov loop contribution from the gluons, respectively. The variation in this quantity, as a function of temperature, is shown in Fig. \ref{qfrac} along both transition lines.

\noindent
Along the LG transition line, the quark fraction is essentially zero for temperatures below 100 MeV (where the interplay between the two crossover transitions is negligible). Above this value it gradually rises (cf. Fig. \ref{qfrac}) as the LG crossover line approaches the de-confinement crossover (cf. Fig. \ref{crit}), thereby introducing an increasing number of quarks in the system.\\
For the chiral transition, the quark fraction starts to increase, quite sharply, at around 100 MeV (cf. Fig. \ref{qfrac}). Below that temperature, the transition is, apparently, a dominantly chiral one, with only a slow change in degrees of freedom. At very low temperatures, however, a slow change in the quark fraction is observed once again. This is because, at these temperatures, quarks can be introduced into the system due to the large chemical potential and due to the quark-suppressing effect of the Polyakov potential disappearing at low temperatures. 
 
\subsection{Susceptibilities and the QCD Phase Diagram}
\label{suscsec}
The thermodynamics of QCD at small values of $\mu_\text{B}/T$ can be obtained by a Taylor expansion of lattice results at $\mu_\text{B} = 0$, in terms of the baryo-chemical potential. Expanding the pressure,
\begin{equation}
P = -\Omega = \frac{T\ln\mathcal{Z}}{V}~,
\label{pressold}
\end{equation}
where $\Omega$ is the grand-canonical potential, $V$ the volume and $\mathcal{Z}$ the grand-canonical partition function,  the corresponding expansion coefficients $c_\text{n}^\text{B}$, or alternatively, the baryon number susceptibilities $\chi_\text{n}^\text{B}$, result as:
\begin{equation}
\frac{\chi_\text{n}^\text{B}}{T^{4-n}} = n! \ c_\text{n}^\text{B} (T) = \frac{\partial^\text{n}(P(T,\mu_\text{B})/T^4)}{\partial(\mu_\text{B}/T)^\text{n}}~. 
\label{susc}
\end{equation}
The behaviour of these coefficients - or susceptibilities - especially, the third-order $\chi_3^{\rm B}$ (skewness) and fourth-order $\chi_4^{\rm B}$ (kurtosis), in and around the phase transitions, are expected to provide stronger signals of criticality, as compared to the second-order $\chi_2^{\rm B}$, because they diverge with a higher power of the correlation length of the order parameter, close to a second-order-type phase transition.

\begin{figure}[t]
\centering
\includegraphics[width=0.35\textwidth, angle = 270]{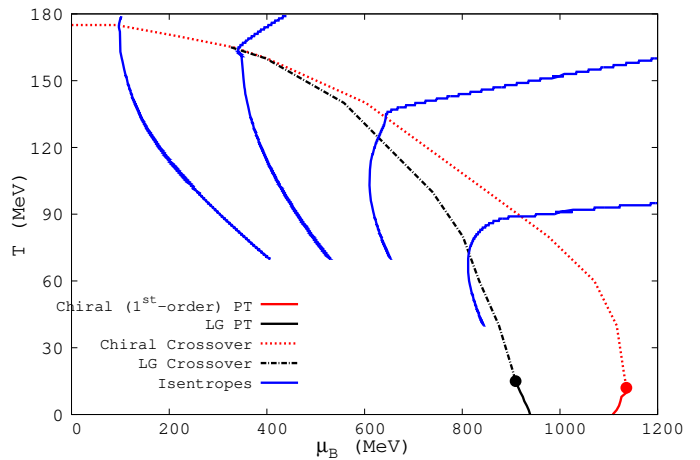}
\caption{[colour online] $T$-$\mu_{\text{B}}$ diagram showing the 1$^{\rm st}$-order liquid-gas (LG) phase transition (bold, black line), the 1$^{\rm st}$-order, chiral phase transition (bold, red line), the LG crossover (dashed, black line), the chiral crossover (dashed, red line), the LG Critical Point (black dot), the chiral Critical Point (red dot) and the isentropes (bold, blue lines) for $S/A$ values 4, 10, 28 and 121, from right to left, respectively.}
\label{crit}
\end{figure}

\noindent
In experiment, usually the normalized ratios $\chi_3^{\rm B}/\chi_2^{\rm B}$ and $\chi_4^{\rm B}/\chi_2^{\rm B}$ are extracted from data, in order to remove the volume dependence of the susceptibilities (N. B.: this does not remove their dependence on volume fluctuations). Before calculating the susceptibilities, it is useful to clearly identify the crossover and first-order phase transition lines of the QCD phase diagram, within this model; as discussed in section \ref{pqf} and as shown in Fig. \ref{crit}.
%A study of the discontinuities of the baryon density ($\rho$) and the chiral condensate ($\sigma$) of the system, as a function of  baryo-chemical potential ($\mu_\text{B}$) and temperature ($T$), allows us to obtain the $T$ and $\mu_\text{B}$ values for the first-order and cross over phase transition curves. As the model is able to describe the nuclear matter ground-state, one thus obtains the liquid-gas phase transition curve as well as the (first-order) chiral phase transition, as shown in Fig. \ref{crit}, with their respective critical end-points.
We observe that both critical end-points occur at a very low temperature. We also observe that the associated crossover lines, while first separated, merge at an intermediate chemical potential $\mu_{\rm B} \approx 400$ MeV.
The figure also shows lines of constant entropy-per-baryon (isentropes) for various values of entropy-per-baryon. The isentropes show a distinct structure, a bending over at the crossover, as the dominant degrees of freedom change from hadrons to quarks. At the junction of the liquid-gas and chiral crossover transitions, the isentropes signal a sharpening of the transition generated by the interplay of the two crossovers.

\noindent
To calculate the susceptibilities, the equations of motion, following from Eqs. (\ref{lagrangian2}) and (\ref{veff}), are solved self-consistently in mean field approximation, by minimising the grand-canonical potential as a function of the baryo-chemical potential and the temperature, as before. Then, the second-, third- and fourth-order derivatives of Eqn. (\ref{susc}) are numerically calculated using a five-point formula. For all temperatures ranging from 15 MeV to 180 MeV, and all baryo-chemical potential values from 0 MeV to 1200 MeV, the results with the previously discussed ratios of susceptibilities are shown in Figs. \ref{tt} and \ref{tt2}.

\noindent
The figures illustrate the effect that the two phase transitions have on the susceptibility values. In the de-confinement phase the susceptibilities have values smaller than 1, as expected for a gas of low-mass fermions. In the region below the LG phase transition (at values of $\mu_\text{B} < 600 \ \text{MeV}$), they are consistently close to 1, since the system is composed of bound hadrons, where a value of 1 for the cumulants of conserved charges is expected. For the region between the crossover transitions from liquid (bound hadrons) to gas (of hadron resonances) and from a hadron resonant gas (HRG) to the QGP, an interplay between the two phase transitions can be observed. This results in the cumulants sometimes having values below 1, or even below zero, and sometimes having values greater than 1, in this intermittent region.

\noindent
In order to give a rough estimate of the susceptibility ratios that could be expected from experiment, one has to define the point in the phase diagram at which the fluctuations are, essentially, frozen out. This point will be different for each beam energy and system size, and in general, is not trivially defined. 
However, it has been found that the measured mean multiplicities of stable hadrons can be nicely described by a thermal fit, with a single value of $T$ and $\mu_\text{B}$, for a specific beam energy. For different beam energies, different $T$ and $\mu_\text{B}$ values are obtained, thus producing the so-called 'Freeze-out Line' \cite{Andronic:2008gu}. By fitting experimental data, the equation of a freeze-out line can be obtained as:

\begin{figure}[t]
\includegraphics[width=0.51\textwidth]{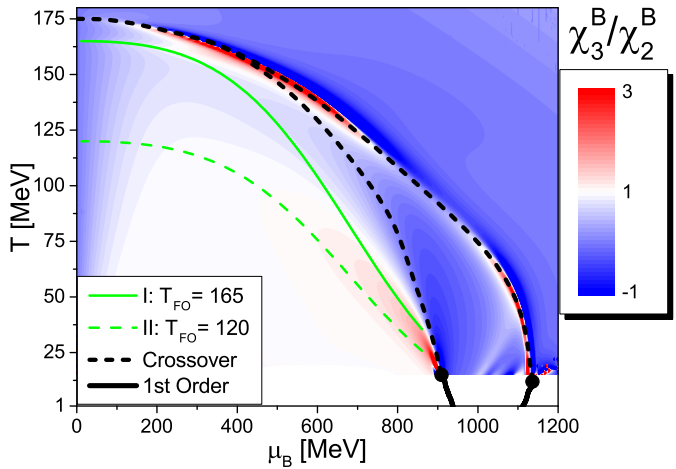}
\caption{[colour online] Susceptibility ratio $\chi_3^{\rm B}/\chi_2^{\rm B}$ as a function of temperature and baryo-chemical Potential. The solid black lines denote the $1^{st}$-order LG and chiral transitions, the dashed black lines denote the crossovers and the green (solid and dashed) lines denote the freeze-out curves for $T_{\rm lim}$ values 165 MeV and 120 MeV, respectively.}
\label{tt}
\end{figure}

\begin{equation}
T \text{ (MeV)} =  \frac{T_{\text{lim}}}{1+ \exp \left[2.60 - \frac{\ln \left(\sqrt{s_{\text{NN}} \text{ (GeV)}}\right)}{0.45}\right]}~,
\label{tfrz}
\end{equation}
where $\mu_\text{B}$ and $\sqrt{s_{\text{NN}}}$ are related as

\begin{equation}
\mu_\text{B} \text{ (MeV)} = \frac{1303}{1+0.286\sqrt{s_{\text{NN}} \text{ (GeV)}}}~;
\label{mfrz}
\end{equation}
with $\sqrt{s_{\text{NN}}}$ being the beam energy in GeV and $T_{\text{lim}}$ being a parameter. Again, one must keep in mind that Eqs. (\ref{tfrz}) and (\ref{mfrz}) represent a mere approximation, and the true freeze-out process is much more complicated than is assumed in this study \cite{Steinheimer:2016cir}. Nevertheless, it is worthwhile to study the behaviour of the normalized cumulants along different possible freeze-out lines.

\noindent
In this study, two different freeze-out lines, obtained by using two different values of $T_{\text{lim}}$ (165 MeV and 120 MeV) in Eqs. (\ref{tfrz}) and (\ref{mfrz}) as shown in Figs. \ref{tt} and \ref{tt2}, are used. Here, the higher value corresponds to the expected latest point of chemical equilibrium while the lower value is closer to the kinetic freeze-out point. For an ideal Boltzmann gas, the susceptibility ratio $\chi_4^{\rm B}/\chi_2^{\rm B}$ along these freeze-out lines can been shown to be equal to 1.

\begin{figure}[t]
\includegraphics[width=0.51\textwidth]{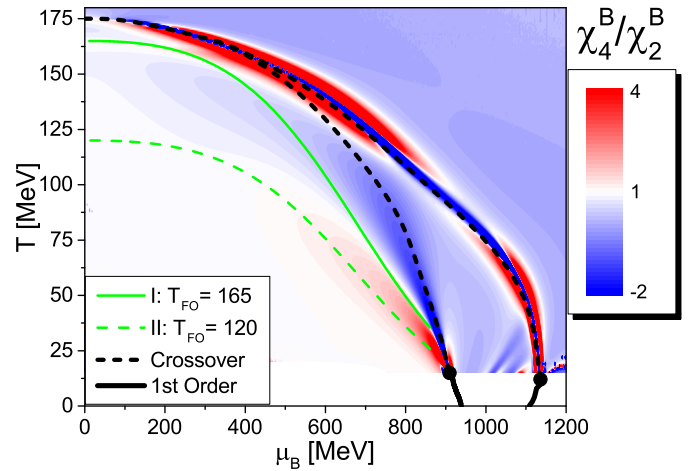}
\caption{[colour online] Same as Fig. \ref{tt} for the susceptibility ratio $\chi_4^{\rm B}/\chi_2^{\rm B}$.}
\label{tt2}
\end{figure}

\begin{figure}[t]
\centering
\includegraphics[width=0.35\textwidth, angle = 270]{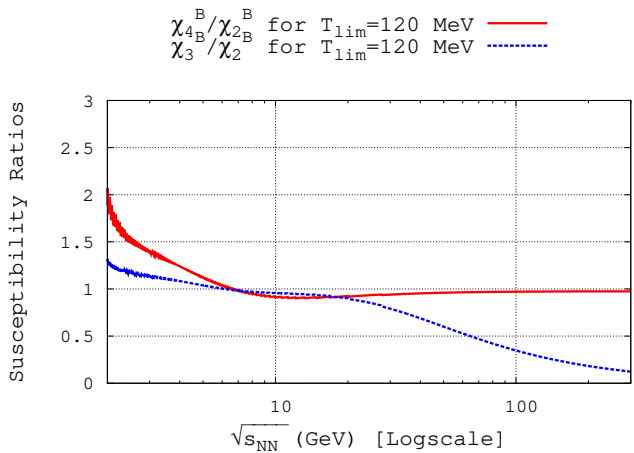}
\caption{[colour online] Susceptibility ratios as function of beam energy along the freeze-out line with $T_{\text{lim}}$ = 120 MeV.}
\label{low1}
\end{figure}

\begin{figure}[t]
\centering
\includegraphics[width=0.35\textwidth, angle = 270]{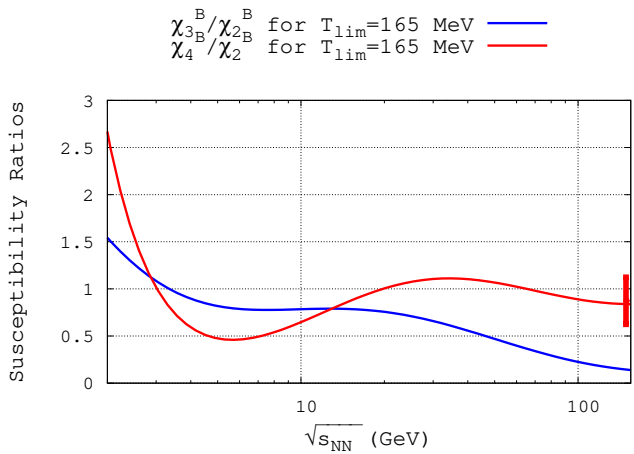}
\caption{[colour online] Same as Fig. \ref{low1} for $T_{\text{lim}}$ = 165 MeV; with the value of $\chi_4^{\rm B}/\chi_2^{\rm B}$ for $\mu_{\rm B} \approx 0$, obtained from lattice data at $T= 150$ MeV \cite{Borsanyi:2013hza}, represented by the thick, red bar.}
\label{high1}
\end{figure}

\noindent
The extracted values of the normalized cumulants are displayed in Figs. \ref{low1} and \ref{high1}, as functions of the beam energy $\sqrt{s_\text{NN}}$.
In the case of the low freeze-out temperature, the measured cumulants essentially resemble those of an ideal HRG, down to beam energies $\sqrt{s_\text{NN}}\le 10$ GeV. Below that energy, the measured susceptibilities actually probe the critical behaviour of the nuclear liquid-gas transition and not that of the QCD chiral transition, as already found in \cite{Fukushima:2014lfa,Vovchenko:2015pya,Vovchenko:2016rkn}. If, however, the higher freeze-out temperature is realized, one can observe a different dependence of the measured cumulants on the beam energy. A peak in the susceptibility ratio is then observed, at a beam energy of $\sqrt{s_\text{NN}}\approx 20$ GeV, due to the steepening of the chiral crossover with respect to chemical potential, at finite $\mu_\text{B}$ (Note: not due to the appearance of a critical point). At lower beam energies, the critical behaviour of the nuclear liquid-gas transition can be observed again.

\noindent
In Fig. \ref{high1} we also compare our results with the value of $\chi_4^{\rm B}/\chi_2^{\rm B}$ which has been extracted from lattice QCD calculations at $\mu_{\rm B}=0$ and $T \approx 150 $ MeV \cite{Borsanyi:2013hza}. One can already see, that the lattice data slightly below $T_{\rm PC}$ still has a significant uncertainty, and a quantitative comparison with our results is difficult for low temperatures.

\noindent
At this point it would be interesting to directly compare our susceptibility ratios with experimental data. As has been shown in, e.g., \cite{Luo:2017faz}; the values of the cumulant ratios extracted from experiment depend strongly on the selected acceptance, as well as the centrality. Furthermore, experiments only measure net-protons; not net-baryons. It is therefore not clear what we should compare our grand canonical values to. One should also keep in mind, that a direct comparison of our grand canonical results with experimental data is not possible due to the many effects discussed in \cite{Bzdak:2012ab,Bzdak:2016qdc,kitazawa2016efficient,Feckova:2015qza,Begun:2004gs,Bzdak:2012an,Gorenstein:2008et,Gorenstein:2011vq,Sangaline:2015bma,Spieles:1996is,Kitazawa:2012at,Asakawa:2000wh,Jeon:2000wg,Steinheimer:2016cir}. The point of our paper is, rather, to discuss the effects of including realistic nuclear matter; in a model with hadron-quark phase transition; on the baryon-number susceptibilities. The eventual comparison of the cumulants to experimental observables has to be determined in a dynamical approach to heavy-ion collisions, which may use our model EoS as an input.

\begin{figure}[t]
\centering
\includegraphics[width=0.35\textwidth, angle = 270]{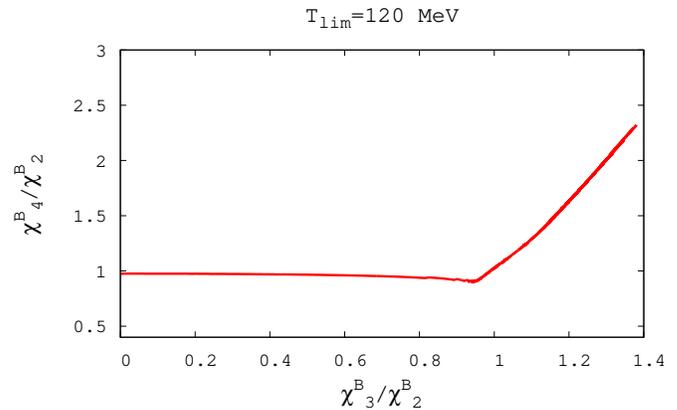}
\caption{[colour online] Susceptibility ratio $\chi_4^B/\chi_2^B$ versus $\chi_3^B/\chi_2^B$,  along the freeze-out line with $T_{\text{lim}}$ = 120 MeV, for beam energies greater than 2 GeV.}
\label{low}
\end{figure}

\begin{figure}[t]
\centering
\includegraphics[width=0.35\textwidth, angle = 270]{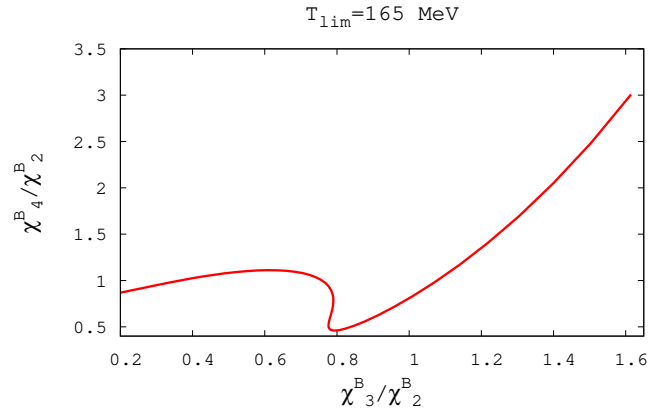}
\caption{[colour online] Same as Fig. \ref{low} for $T_{\text{lim}}$ = 165 MeV.}
\label{high}
\end{figure}

\noindent
It was pointed out in \cite{Chen:2016sxn} that, given a critical point of a particular universality class (and only one critical point!), the dependence of the normalized cumulants, as functions of one another, should show a particular universal banana-type shape.

\noindent
Figs. \ref{low} and \ref{high} show the shapes obtained from the parity-doublet model calculations. Due to the fact that this model actually has two separate transitions, which are difficult to disentangle, the resulting shapes do not resemble a banana, but are more complicated. In general, when there is an interplay between two phase transitions the relationship between the skewness and the kurtosis is affected by the remnants of the crossover regions related to both the LG and chiral transition, as shown in Fig. \ref{high}. Even for the $T_{\text{lim}}$ = 120 MeV freeze-out line (cf. Figs. \ref{tt} and \ref{tt2}), the aforementioned interactions, for $\sqrt{s_\text{NN}} \ge 2$ GeV, give results considerably different from those which are obtained using universality arguments (cf. Fig. \ref{low}), as only the liquid-gas transition is observed.  

\noindent
An important result of this work is the strong dependence of the range of values for the ratios, at large beam energies, on the choice of the freeze-out point. Since both transitions can have an impact on the observed cumulant ratios, it is therefore important to understand the point of origin, during the system's evolution, of the measured fluctuations, a problem which cannot be solved within the bounds of the present model, as it requires a dynamical description of the nuclear collisions, including the propagation of critical fluctuations.

\section{Conclusion}
We have presented an improved version of the hadronic, three-flavour, parity-doublet model including a de-confinement transition to quarks and gluons. The main modification is the inclusion of a six-point interaction term, which significantly improves the nuclear matter saturation properties of the model. With this, we have constructed a model which gives a satisfactory description of nuclear matter, as well as qualitatively describes lattice QCD thermodynamics at $\mu_{\rm B}=0$.\\
We have employed the model to study the interplay between the nuclear liquid-gas transition and the chiral transition at large temperatures. We find that this interplay does have an effect on the equation of state and the extracted susceptibilities in a significant range of the phase diagram. This means that the influence of dense nuclear matter on the phase structure, even at large temperatures and moderate chemical potentials, cannot be neglected.\\
Furthermore, we have studied the beam energy dependence of the normalized cumulants from our model for different freeze-out conditions. Again, we observe a strong influence of nuclear matter interactions on the observed fluctuations, particularly for low beam energies.\\
Our work highlights the fundamental importance of consistently including the properties of interacting nuclear matter in an effective model of the QCD Equation of State for interpreting  experimental data of particle fluctuations in heavy-ion collisions.\\

\section{Acknowledgements}
This work was supported by BMBF. The computational resources were provided by the LOEWE Frankfurt Centre for Scientific Computing (LOEWE-CSC).

%%%%%%%%%%%%%%%%%%%%%%%%%%%%%%%%%%%%%%%%%%%%%%%%%%%%%%%%%%%%%%%%%%%%%%%%%%%%%%%
%\begin{thebibliography}{100}

\bibstyle{apsrev4-1}
\bibliography{bibnew.bib}

% Save this file and include it in your paper as the bibliography
% or cut and paste directly into your LaTeX

%\end{thebibliography}%%%%%%%%%%%%%%%%%%%%%%%%%%%%%%%%%%%%%%%%%%%%%%%%%%%%%%%%%%%%%%%%%%%%%%%%%%%%%%% 

\end{document}